\documentclass[twocolumn,aps,prl,superscriptaddress,showpacs,amsmath,amssymb,floats]{revtex4}
\usepackage{amssymb}
\usepackage{amsmath}
\usepackage{graphicx}
\usepackage{CJK}
\usepackage{keyval}
\usepackage{dcolumn}
\usepackage{bm}
\usepackage{color}
\usepackage{txfonts}
\usepackage[]{sidecap}

\begin{document}

\begin{CJK*}{UTF8}{gbsn}

\title{Highly anisotropic and two-fold symmetric superconducting gap in nematically ordered FeSe$_{0.93}$S$_{0.07}$}

\author{H. C. Xu}\author{X. H. Niu}\author{D. F. Xu}\author{J. Jiang}\author{Q. Yao}
\affiliation{State Key Laboratory of Surface Physics, Department of Physics, and Advanced Materials Laboratory,
Fudan University, Shanghai 200433, People's Republic of China}

\author{M. Abdel-Hafiez}
\affiliation{Center for High Pressure Science and Technology Advanced Research, Shanghai 201203, China}
\affiliation{Faculty of science, Physics department, Fayoum University, 63514-Fayoum, Egypt}

\author{D. A. Chareev}
\affiliation{Institute of Experimental Mineralogy, Russian Academy of Sciences, 142432 Chernogolovka, Moscow District, Russia}

\author{A. N. Vasiliev}
\affiliation{Low Temperature Physics and Superconductivity Department, M.V. Lomonosov Moscow State University, 119991 Moscow, Russia}

\author{R. Peng}\email{pengrui@fudan.edu.cn}
\affiliation{State Key Laboratory of Surface Physics, Department of Physics, and Advanced Materials Laboratory,
Fudan University, Shanghai 200433, People's Republic of China}

\author{D. L. Feng}\email{dlfeng@fudan.edu.cn}
\affiliation{State Key Laboratory of Surface Physics, Department of Physics, and Advanced Materials Laboratory,
Fudan University, Shanghai 200433, People's Republic of China}

\date{\today}
\begin{abstract}
FeSe exhibits a novel ground state in which superconductivity coexists with a nematic order in the absence of any long-range magnetic order.
Here we report an angle-resolved photoemission study on the superconducting gap structure in the nematic state of FeSe$_{0.93}$S$_{0.07}$, without the complication caused by Fermi surface reconstruction induced by magnetic order.  We found that
the superconducting gap shows a pronounced 2-fold anisotropy around  the elliptical hole pocket near the Z point of the Brillouin zone, with gap minima at the endpoints of its major axis, while no detectable gap was observed around the zone center and zone corner.
The large anisotropy and nodal gap distribution demonstrate the substantial effects of  the nematicity on the superconductivity, and thus put strong constraints on the  current theories.
%These results provide the foundation for constructing a more comprehensive theory of this important family of Fe-based superconductors.

\end{abstract}

\pacs{74.70.Xa, 74.25.Jb, 74.20.Mn}

\maketitle

\end{CJK*}

The pairing mechanism underlying unconventional superconductivity is often related to the quantum fluctuations of nearby orders. In most Fe-based superconductors, both magnetic and nematic orders appear simultaneously near the superconducting state. Accordingly, both  spin-fluctuation-mediated and orbital-fluctuation-mediated superconducting pairing mechanisms have been proposed \cite{Hirschfeld-review,GlasbrennerNP2015,YuRPRL2015,WangFNP2015,KontaniOrbital2010}.
Although intense experimental studies have been conducted \cite{YZhangKFeSe2011,DFLiuFeSeNC2012,MXuPRB2012,BZengPRB2011,WYuPRB2011,JTParkPRL2011,PengRPRL2014,FanQNP2015}, the exact pairing mechanism of Fe-based superconductors is still under heated debate.

FeSe is a unique material with a novel superconducting state. Orbital order develops in the nematic state of FeSe without breaking the translational symmetry as shown by angle resolved photoemission spectroscopy (ARPES) studies \cite{WatsonFeSe2015,YZhangFeSe2015}. The superconductivity coexists with the nematic order without any long range magnetic order \cite{McQueenPRL2009}, thus disentangling the magnetic and orbital orders.
Moreover, recent results suggest that FeSe is a quantum paramagnet \cite{WangFNP2015} with coexisting N\'{e}el and stripe antiferromagnetic interactions \cite{WangQSNeutronNM2016,WangQSNeutron2016}. The novel ground state in FeSe provides a fresh perspective for studying the effect of nematic order on the superconducting gap structure in the absence of the Fermi surface reconstruction induced by magnetic order,
which helps to reveal the roles of spin and orbital degrees of freedom in unconventional superconductivity.
%The superconducting gap structure provides critical information on superconducting pairing.
A nodeless superconducting gap structure in FeSe was suggested by previous reports on specific heat \cite{LinPRB2011}, Andreev reflection spectroscopy \cite{Chareev2013}, and thermal conductivity measurements \cite{DongJKPRB2009}. In contrast, scanning tunnelling spectroscopy (STS) studies on FeSe films \cite{SongSTM2011} and transport measurements on bulk FeSe/FeSe$_{1-x}$S$_{x}$ crystals with improved quality \cite{KasaharaSTM2014,MoorePRB2015} all demonstrate a nodal gap structure.
However, due to the low $T_c$ and small gap size of FeSe/FeSe$_{1-x}$S$_{x}$ single crystals, the gap distribution in momentum-space is still unknown.

In this work, we studied  the superconducting gap structure of high-quality FeSe$_{0.93}$S$_{0.07}$ single crystals ($T_c$ = 10~K) with high resolution ARPES \cite{HafiezFeSeS2015}. At 6.3~K, both the nematic electronic structure  and the superconducting gap are observed. The gap amplitude at the hole pocket is 2.5~meV , similar to that measured by STS \cite{MoorePRB2015}.
The superconducting gap shows 2-fold anisotropy around the Z point, and is undetectable around the hole Fermi surface near the zone center and the electron pockets at the zone corners. We find that the unique gap structure observed here cannot be resonably fitted by most known theoretical gap structures and their simple combinations, which suggest that the effects of nematicity on the superconductivity are substantial.

\begin{figure*}[tb]
\includegraphics[width=17cm]{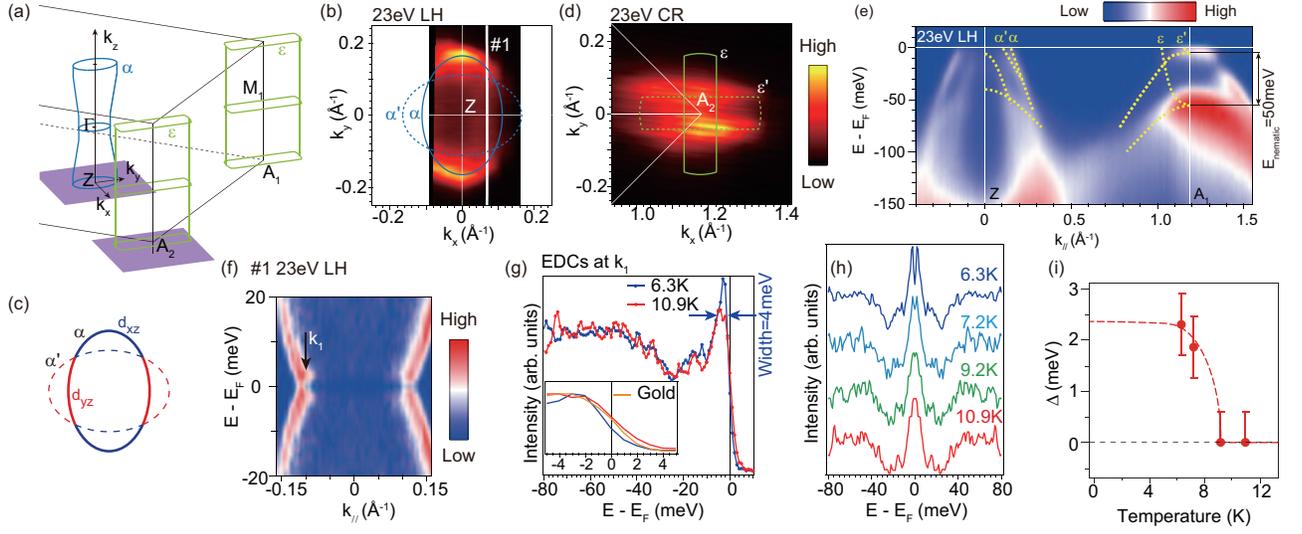}
\caption{(color online).
(a) Fermi suface topology in the 2-Fe Brillouin zone of FeSe in the nematic state according to Refs.~\cite{WatsonFeSe2015}.
(b) Fermi surface mapping around the Z point with linear-horizontal (LH) polarized photons. The corresponding momenta are indicated by the purple square in panel (a).
(c) Orbital contents of pocket $\alpha$ (solid ellipse) and its counterpart from another twin domain ($\alpha'$, dashed ellipes) according to ref. \cite{WatsonFeSe2015}.
(d) Same as (b) but around the A$_2$ point with circular-right (CR) polarized photons.% The solid curves illustrate the Fermi surfaces, while the dashed curves illustrate those from twin domains. The inset in panel (b) shows the orbital contents of pockets $\alpha$ and $\alpha'$ in FeSe according to ref. \cite{WatsonFeSe2015}.
(e) Photoemission intensity along Z-A$_1$. The band splitting $E_{nematic}$ due to nematic orbital ordering is indicated.
(f) Symmetrized photoemission intensity along cut~\#1 as indicated in panel (c).
(g) Energy distribution curves (EDCs) above and below $T_c$ at the momentum $k_1$ in panel (f). The width of the superconducting quasiparticle peak is indicated. The inset shows the leading edge shift near the Fermi energy. % The full width at half maximum ($\Gamma_{\mathrm{FWHM}}$) of the superconducting quasiparticle peak fitted by a Lorentz peak and a Shirley background.
(h) Temperature dependent symmetrized EDCs at momentum $k_1$.
(i) Superconducting gap size as a function of temperature fits to the Bardeen-Cooper-Schrieffer formula.
}
\label{FSGap}
\end{figure*}

FeSe$_{0.93}$S$_{0.07}$ single crystals were grown using AlCl$_3$/KCl flux in a temperature gradient (from 400~$^{\circ}$C to $\sim$50~$^{\circ}$C) for 45 days \cite{HafiezFeSeS2015,ChareevGrowth2013}. The ARPES measurements were conducted at the I05 beamline at the Diamond Light Source.
The data were taken at the temperature of 6.3~K unless otherwise specified. The single crystals were cleaved in-situ and measured under ultra-high vacuum of 1$\times$10$^{-10}$~mbar. For data collection with 23~eV (37~eV) photons, the energy resolution was 3~meV (5~meV). Empirically, this allows resolving a superconducting gap of 0.6~meV (1~meV).

%The Fermi surface structure of FeSe$_{0.93}$S$_{0.07}$ is similar to that of FeSe/FeSe$_{1-x}$S$_{x}$ reported previously \cite{WatsonFeSe2015,WatsonFeSeS2015}.
Figure~\ref{FSGap}(a) illustrates the Fermi surfaces of FeSe/FeSe$_{1-x}$S$_{x}$ in the nematic state, which consist of one hole pocket at the zone center and one electron pocket at the zone corner of the 2-Fe Brillouin zone \cite{WatsonFeSe2015,MaletzCalc2014}. There is another electron pocket ($\delta$) around A$_1$/A$_2$ according to the calculation and quantum oscillation measurements\cite{WatsonFeSe2015,MaletzCalc2014}, but it has not been detected by ARPES probably due to its small matrix element \cite{WatsonFeSe2015}.
%Figures~\ref{FSGap}(c)-(d) show the photoemission intensity map of FeSe$_{0.93}$S$_{0.07}$  at 6.3~K with 23~eV horizontally polarized photons, which corresponds to the Z-A$_1$-A$_2$ plane, assuming the inner potential to be 12~eV.
In our data, the elliptical hole pocket $\alpha$ around the Z point is resolved in Fig.~\ref{FSGap}(b). Another elliptical hole pocket $\alpha'$, with weaker spectral intensity, is contributed by 90~$^{\circ}$-rotated twin domains. The distribution of spectral weight is nearly identical to that of FeSe under the same experimental geometry with linear horizontal polarized photons \cite{WatsonFeSe2015}, indicating the similar orbital contents on the $\alpha$ pocket [Fig.~\ref{FSGap}(c)].
The electron pocket $\varepsilon$ around the zone corner (A$_1$/A$_2$) is elongated [Fig.~\ref{FSGap}(d)].
Figure~\ref{FSGap}(e) shows the photoemission spectra along Z-A$_1$, clearly indicating the splitting of 50~meV between bands $\varepsilon$ and $\varepsilon'$  due to the nematic order \cite{WatsonFeSe2015}.

%The Superconducting gap is resolved on band $\alpha$ at 6.3~K.
%As shown by the photoemission spectra in Fig.~\ref{FSGap}(f),
At 6.3~K, band $\alpha$ shows a back-bending dispersion and gap opening [Fig.~\ref{FSGap}(f)], which are the hallmarks of Bogliubov quasiparticles. Sharp quasiparticle peaks are observed at the Fermi crossings of band $\alpha$ at 10.9~K, and become even sharper at 6.3~K [Fig.~\ref{FSGap}(g)], indicating the high quality of the FeSe$_{0.93}$S$_{0.07}$ single crystals. From 10.9~K to 6.3~K, the leading edge shifts below the Fermi energy [inset of Fig.~\ref{FSGap}(g)], indicating the opening of the superconducting gap. The superconducting gap size determined by the symmetrized energy distribution curves (EDCs) is around 2.5~meV at 6.3K [Fig.~\ref{FSGap}(h)], which decreases with increasing temperature and eventually closes around 9.2~K following the Bardeen-Cooper-Schrieffer formula [Fig.~\ref{FSGap}(i)], consistent with the $T_c$ of 10~K measured by transport experiments \cite{HafiezFeSeS2015}. %The gap versus temperature curve follows the Bardeen-Cooper-Schrieffer formula, consistent with the superconducting nature.% These results indicates the coexistence of nematicity and superconductivity in FeSe$_{0.93}$S$_{0.07}$ from the electronic structure point of view.

\begin{figure}
\includegraphics[width=8.6cm]{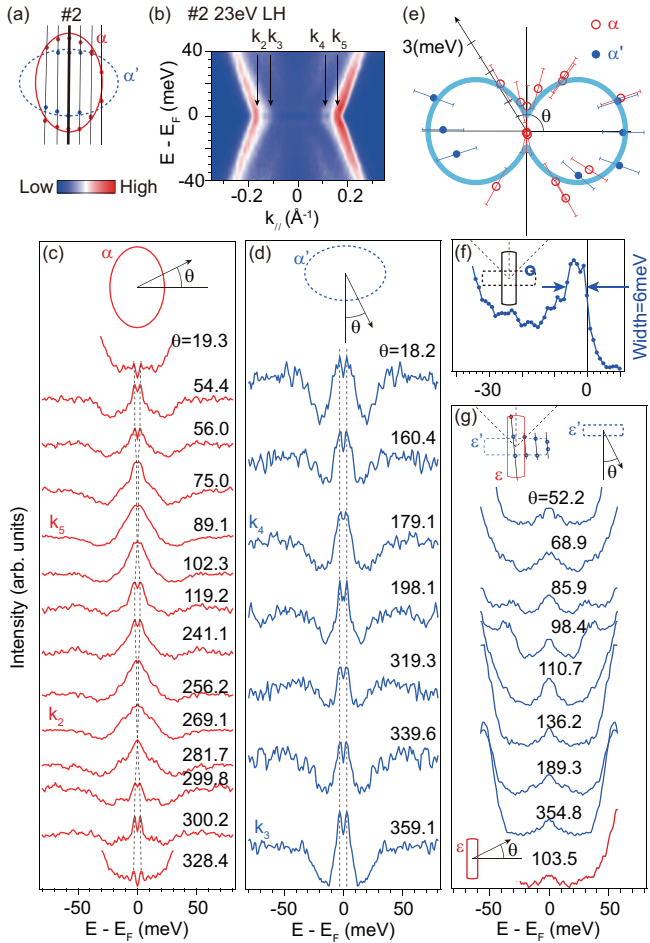}
\caption{(color online).
(a) Illustration of hole pockets $\alpha$ and $\alpha'$ around the Z point.
(b) Symmetrized photoemission intensity along the cut~\#2 as indicated in panel (a).
(c) Symmetrized EDCs on the pocket $\alpha$. The insets define the in-plane angle $\theta$.
(d) Symmetrized EDCs on the pocket $\alpha'$.
(e) Polar plot of the superconducting gap as a function of $\theta$ along the pockets $\alpha$ and $\alpha'$.
(f) EDC at the pocket $\varepsilon'$ indicated by the blue circle.
(g) Symmetrized EDCs on the electron pockets $\varepsilon$ and $\varepsilon$' around the A$_1$ point. The pockets $\varepsilon$' and $\varepsilon$ are from different twin domains.
}
\label{gapZ}
\end{figure}

The momentum distribution of the superconducting gap on the hole pockets $\alpha$ and $\alpha'$ has been studied along the parallel momentum cuts in Fig.~\ref{gapZ}(a) with 23~eV photons.
In the symmetrized photoemission intensity along cut \#2,  $\alpha'$ is gapped at momenta $k_3$ and $k_4$, whereas $\alpha$ crosses the Fermi level at momenta $k_2$ and $k_5$ without any observable gap opening [Fig.~\ref{gapZ}(b)], indicating distinct gap sizes between the momenta near the major axis endpoints of pocket $\alpha$ and those near the minor axis endpoints of pocket $\alpha'$.
As shown by the symmetrized EDCs in Fig.~\ref{gapZ}(c), the superconducting gap is reduced around the major axis endpoints of the elliptical Fermi surface $\alpha$ ($\theta\simeq$~90~$^{\circ}$ and 270~$^{\circ}$). Around the minor axis endpoints of pocket $\alpha'$, the gap size remains constant [Fig.~\ref{gapZ}(d)]. By the empirical fitting to a superconducting spectral function \cite{ZhangYNP2012}, the sizes of superconducting gap as a function of polar angle $\theta$ are summarized in one single polar plot [Fig.~\ref{gapZ}(e)], noting that the $\alpha$ and $\alpha'$ are identical bands from twin domains. The superconducting gap on band $\alpha$ shows anisotropy with 2-fold symmetry. The gap size decreases from about 2.5~meV at the minor axis endpoints of the ellipse, to less than 0.6~meV around the major axis endpoints, which is at the experimental resolution limit.
% (around 0.6~meV estimated from the energy resolution of 3~meV with photon energy 23~eV)

%Around the zone corner, superconducting properties has been investigated along momentum slices across the electron pocket $\varepsilon$ with photon energy 23eV.
The photoemission spectra at the Fermi crossing of band $\varepsilon$' show sharp quasiparticle peaks in the superconducting state [Fig.~\ref{gapZ}(f)].
However, no superconducting gap is detected along the electron pockets $\varepsilon$ or $\varepsilon$' [Fig.~\ref{gapZ}(g)]. The absence of a superconducting gap at these momenta indicates nodes or a small gap size below the experimental resolution limit. % The resolution limit could be around 0.6~meV, which is estimated by the energy resolution of 3~meV while collecting the data with photon energy 23~eV.

\begin{figure}[tb]
\includegraphics[width=8.6cm]{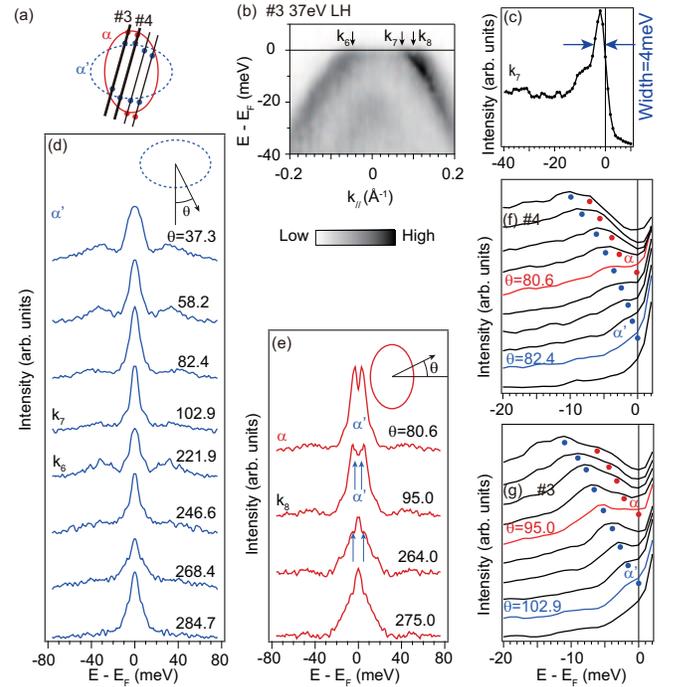}
\caption{(color online).
(a) Illustration of pockets $\alpha$ and $\alpha'$ around the $\Gamma$ point.
%The sketch at the right defines the in-plane angle $\theta$.
(b) Photoemission intensity along the cut \#3 as indicated in panel (a).
(c) The EDC at momentum $k_7$.
(d) Symmetrized EDCs on the pocket $\alpha'$. The inset defines the in-plane angle $\theta$.
(e) Symmetrized EDCs on the pocket $\alpha$. Quasiparticle peaks from band $\alpha$' are indicated.
(f) EDCs divided by the resolution-convolved Fermi-Dirac function near the upper Fermi crossings of cut \#4 in panel (a).
(g) Same as panel (f) but near the upper Fermi crossings of cut \#3.
}
\label{GapG}
\end{figure}

%The superconducting gap is also investigated  around $\Gamma$ point with 37~eV photons.
Figure~\ref{GapG}(a) illustrates the photoemission cuts through bands $\alpha$ and $\alpha'$ around the $\Gamma$ point with 37~eV photons. The bands $\alpha$ and $\alpha'$ are resolved along cut \#3 [Fig.~\ref{GapG}(b)], showing sharp quasiparticle peaks at the Fermi crossings [Fig.~\ref{GapG}(c)]. Along the elliptical Fermi surface $\alpha'$, the symmetrized EDCs show no detectable superconducting gap [Fig.~\ref{GapG}(d)], indicating nodes or a small gap size below the experimental resolution limit.
% (around 1~meV estimated from the energy resolution of 5~meV at 37~eV photon energy).
For band $\alpha$, the Fermi crossings with polar angles 264.0~$^{\circ}$ and 275.0~$^{\circ}$ show no observable gap either [Fig.~\ref{GapG}(e)]. The quasiparticle peaks at $\sim\pm4$~meV for $\theta=$ 80.6~$^{\circ}$ and 95.0~$^{\circ}$ are contributed by band $\alpha'$, which gives  false signatures of gap opening in  Fig.~\ref{GapG}(e). Actually as shown in Figs.~\ref{GapG}(f) and \ref{GapG}(g), the EDCs divided by the resolution-convolved Fermi-Dirac function are flat within 2-3~meV of the Fermi crossings of band $\alpha$, indicating no detectable gap opening.
As shown in the Supplementary Material \cite{SM},
the  gap amplitude decreases from Z to  $\Gamma$ untill it diminishes, which  is intriguingly opposite to  those observed in BaFe$_2$(As$_{1-x}$P$_x$)$_2$ and Ba$_{1-x}$K$_x$Fe$_2$As$_2$ \cite{ZhangYNP2012,ZhangYPRL2010}, where the gap of the $\alpha$ band decreases from $\Gamma$ to Z.

%was explained by possible mixing of $d_{3z^2-r^2}$ orbital at the Fermi energy \cite{Laad2009,WangF2010,Suzuki2011,SuY2011}. Similar picture could apply in our case, because the FeSe show electronic structure with relatively strong 3-dimensional character \cite{WatsonFeSe2015}.

\begin{figure}[tb]
\includegraphics[width=8.6cm]{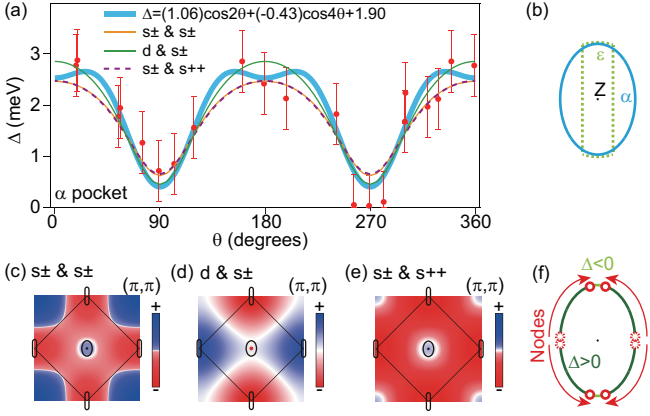}
\caption{(color online).
%(a) The Fe-layer lattice structure in the nematic state.
%The projected 1-Fe Brillouin zone and Fermi surface in nematic state of FeSe. %The nodal lines of $J_1$ s-wave and $J_2$ s-wave are indicated by red and blue dashed lines, respectively.
% The orbital composition of Fermi surfaces are reproduced according to previous reports \cite{WatsonFeSe2015,MukherjeePRL2015}.
(a) Angular dependence of the superconducting gap on the $\alpha$ pocket and fitting results by cosine series, $\Delta_{s\pm,s\pm}$, $\Delta_{d,s\pm}$, and $\Delta_{s\pm,s\mathrm{++}}$. %$\Delta=\Delta_0+\Delta_1\cos{2\theta}+\Delta_2\cos{4\theta}$, $\Delta=\Delta_1\cos{k_x}\cos{k_y}+\Delta_2(\cos{k_x}+\cos{k_y})/2$ (i.e. $s\pm$ \& $s\pm$), $\Delta=\Delta_1(\cos{k_x}-\cos{k_y})/2+\Delta_2(\cos{k_x}+\cos{k_y})/2$ (i.e. $d$ \& $s\pm$), and $\Delta=\Delta_2(\cos{k_x}+\cos{k_y})/2+\Delta_s$ (i.e. $s\pm$ \& $s$++).
(b) Overlapping of the pockets $\varepsilon$ (dashed curves) and $\alpha$ (solid curves) in the case of antiferromagnetic folding.
(c-e) The gap forms obtained by the fitting results in panel (a).
(f) The gap structure on the pocket $\alpha$ according to the theory of Ref.~\cite{KangNodes2014}.
}
\label{GapFit}
\end{figure}

%Our data resolve the momentum location of the nodes and the gap with size smaller than 0.6~meV or 1~meV at 6.3~K.
%The absence of detectable superconducting gap at various momenta of Fermi surface $\alpha$ and $\varepsilon$ could be attributed to the small gap size and limited energy resolution.
%Future measurements with higher resolution and lower temperature are required to distinguish the small gap and gap nodes.
%However, superconducting gap nodes should only exist at a small portion of the Fermi surface, because the V-shaped dI/dV curve reaches zero residual density of states  at 0.4~K in previous STM reports \cite{SongSTM2011,MoorePRB2015}. A relatively high temperature reduce the superconducting gap sizenodes as shown in previous STM studies \cite{SongSTM2011}, which explains the absence of gap at various momenta in our data. Our results reveal the gap nodes at 6.3~K, while the nodes at 0.4~K should exist within these momenta.
%More importantly, the distribution of gap size in the momentum space reveals valuable information on the superconducting pairing, especially the novel 2-fold symmetric gap at the Fermi surface $\alpha$.

Our results confine the nodes to the vicinity of the two endpoints on the elliptical $\alpha$ pocket around Z, the  $\alpha$ pocket around $\Gamma$, and the electron pockets.
Considering that the STS spectra on superconducting FeSe$_{1-x}$S$_{x}$ \cite{MoorePRB2015} show finite but small density of states at the Fermi energy at 0.4~K, the nodes can only occur on a small portion of the Fermi surface, while most of the momenta without a detectable gap in our data must exhibit a finite gap at much lower temperatures. Though the precise positions of nodes in these regions will have to be determined with better resolution in future studies, the momentum dependent gap structure,
especially the large gap anisotropy at the $\alpha$ pocket showing remarkable component of $\cos{2\theta}$ with 2-fold symmetry [Fig.~\ref{GapFit}(a)],
%specifically the large gap anisotropy at the $\alpha$ pocket and the undetectable gap size around the A$_1$/A$_2$ and $\Gamma$ points,
put constraints on current theories of superconductivity in FeSe.
%Fitting the gap anisotropy at the $\alpha$ pocket gives remarkable component of $\cos{2\theta}$ with 2-fold symmetry [Fig.~\ref{GapFit}(a)].
The the observed gap structure can be used to scrutinize four types of current scenarios:

First, in the case of superconductivity with dominant $s$++ pairing mediated by orbital fluctuations, the gap form is nearly isotropic and nodeless \cite{KontaniOrbital2010}. The large anisotropy and nodal behavior of the gap in FeSe suggest that the superconducting pairing in FeSe is not mediated by pure orbital fluctuations.

%The gap anisotropy may come from the elliptical Fermi surface $\alpha$ in the gap form of spin-fluctuation mediated s+- pairing.
Second, in $s\pm$ pairing mediated by magnetic interactions, the sign-changing gap form may lead to gap anisotropy and nodes \cite{ScalapinoReview2012,JPHu2012}. Since both N\'{e}el and stripe spin-fluctuations exist in FeSe \cite{WangQSNeutron2016}, if the $s\pm$ superconducting pairing were generated either by the ($\pi,\pi$) interaction with gap form $\cos{k_x}\cos{k_y}$, or by the ($\pi,0$) interaction with gap form $(\cos{k_x}+\cos{k_y})/2$ \cite{JPHu2012}, the anisotropy of the superconducting gap on the elliptical $\alpha$ pocket would be 3\% and 6\% for these two gap forms, respectively. These cannot account for the large anisotropy of at least 78\% observed in our data.
%For the gap forms from either $J_1$ or $J_2$ interaction, the Fermi surface $\alpha$ is too small to produce a large anisotropy of superconducting gap as observed in our data [Fig.~\ref{GapFit}(b)].

If there were static stripe antiferromagnetic order with wave vector ($\pi$, 0), the electron pockets would have been folded to the zone center and intersect with the $\alpha$ pocket around the major axis endpoints in FeSe [Fig.~\ref{GapFit}(b)]. Theory suggests that gap nodes would emerge at the reconstructed Fermi surfaces, given a large value of antiferromagnetic order parameter \cite{MaitiPRB2012}. However, FeSe shows no static magnetic order, thus its gap anisotropy cannot be explained by this scenario. %Further calculations are required to check whether the possible dynamic folding in FeSe would lead to the gap anisotropy and nodes observed in our data.
%The antiferromagnetic order could induce Fermi surface reconstruction and gap nodes \cite{MaitiPRB2012}, but it requires coexisting static long range magnetic order and large antiferromagnetic order parameter, which is absent in  FeSe.
%The stripe antiferromagnetic order \cite{WangQSNeutron2016} folds the electron pockets to the zone center and mix it with the elliptical pocket $\alpha$ around the major axis endpoints with opposite superconducting pairing phase [Fig.~\ref{GapFit}(a)]. thereory suggest gap nodes would be induced only when the static magnetic order is very strong. but the strong order is absent in FeSe. %Theory of antiferromagnetism induced gap nodes requires coexisting static long range magnetic order and large antiferromagnetic order parameter \cite{MaitiPRB2012}, which is absent in  FeSe.
%Even if the antiferromagnetic spin fluctuations with ($\pi$, 0) wave vector \cite{WangQSNeutron2016} dynamically mix the gaps around Z and A with opposite phases in the $(cosk_x+cosk_y)/2$ gap form [Fig.~\ref{GapFit}(a)], the sign change due to the spin flip during the antiferromagnetic folding would compensate the phase difference and lead to nodeless gap structure in contrast to experimental observations \cite{ParkerPRB2009}.

Third,
a composite form of superconducting pairing may arise from the quantum paramagnet ground state with N\'{e}el and stripe spin fluctuations \cite{WangFNP2015,WangQSNeutron2016}.
%Considering the coexistence of the two spin fluctuations of both $J_1$ and $J_2$ interactions,
In Fig.~\ref{GapFit}(a), we fit the gap anisotropy of the $\alpha$ pocket by \cite{SongSTM2011}
\[\Delta_{s\pm,s\pm}=\Delta_1\cos{k_x}\cos{k_y}+\Delta_2(\cos{k_x}+\cos{k_y})/2, \]
which gives superconducting gap sizes $\Delta_1$=-58.2$\pm8.8$~meV and $\Delta_2$=62.2$\pm9.2$~meV for $s\pm$ pairing mediated by the two kinds of spin fluctuations.
Moreover, the combination of N\'{e}el spin fluctuation mediated $d$-wave pairing and stripe spin fluctuation mediated $s\pm$ pairing with the gap form
\[\Delta_{d,s\pm}=\Delta_d(\cos{k_x}-\cos{k_y})/2+\Delta_2(\cos{k_x}+\cos{k_y})/2,\]
also gives good fitting with $\Delta_1$=30.3$\pm2.8$~meV and $\Delta_2$=2.24$\pm0.09$~meV [Fig.~\ref{GapFit}(a)].
% which are unphysical considering the low $T_c$ of FeSe and would lead to large gap at zone corner.
Alternatively, by combining the spin-fluctuation-mediated $s\pm$ pairing and orbital-fluctuation-mediated $s$++ pairing \cite{KontaniOrbital2010}, the gap anisotropy at pocket $\alpha$ can be fitted by
\[\Delta_{s\pm,s\mathrm{++}}=\Delta_2(\cos{k_x}+\cos{k_y})/2+\Delta_s,\]
with $\Delta_2$=32.8$\pm4.8$~meV and $\Delta_s$=-28.7$\pm4.4$~meV for $s\pm$ and $s$++ pairing, respectively [Fig.~\ref{GapFit}(a)].
%The energy scales of superconducting pairing from the above fits are much larger than the superconducting gap of around 2.5~meV.
All three fittings contain gap amplitudes over 30~meV, which are nonphysical compared with the low $T_c$ of FeSe. Moreover, the obtained gap forms would give a large gap at the $\varepsilon$ pocket [Figs.~\ref{GapFit}(c)-\ref{GapFit}(e)], in contrast to the undetectable superconducting gap in our data.
%Moreover, the obtained gap forms in Figs.~\ref{GapFit}(c)-\ref{GapFit}(e) would give rise to a large gap of over 30~meV at the $\varepsilon$ pocket around the A$_1$/A$_2$ point, in contrast to the undetectable superconducting gap in our data.
Therefore, these simple combinations of gap forms cannot account for the large gap anisotropy on pocket $\alpha$.% The effect of nematic state on the symmetry of superconducting gap form should be considered.

Fourth, we consider an orbital-dependent superconducting pairing symmetry. The orbital anti-phase pairing cannot explain the gap anisotropy on pocket $\alpha$ \cite{YinZPOrbital,LuPRB2012}, because the orbital composition changes around $\theta=\pm45^{\circ}$ and $\pm135^{\circ}$ [Fig.~\ref{FSGap}(c)], rather than at the major axis endpoints where the gap minima appear. On the pocket $\alpha$, the Fermi surface sections showing gap minima and gap maxima coincide with those with $d_{xz}$ and $d_{yz}$ orbital characters, respectively, indicating that the orbital ordering may lead to weaker superconducting pairing of $d_{xz}$ orbital than that of $d_{yz}$ orbital. Alternatively, it was shown that the orbital ordering may mix different pairing symmetries and give rise to pairs of accidental nodes \cite{KangNodes2014}. The positions of the nodes depend on the splitting between $d_{xz}$ and $d_{yz}$ orbital, which was set to 80~meV in the theory, while it is 50~meV in FeSe$_{0.93}$S$_{0.07}$ [Fig.~\ref{FSGap}(e)].
%The orbital order and the multiple spin fluctuations in superconducting FeSe \cite{WangFNP2015,WangQSNeutron2016} meet the prerequisite of this theory.
%Further calculations are required to study how the orbital ordering in FeSe$_{0.93}$S$_{0.07}$ would move the nodes.
In this scenario, if a pair of nodes are located very close to a major-axis endpoint of the $\alpha$ pocket due to a strong nematic order [Fig.~\ref{GapFit}(f)], the gap would exhibit just one minimum at each endpoint in the data due to the limited momentum resolution, which would be consistent with our findings.

In summary, we have revealed the superconducting gap structure of  FeSe$_{0.93}$S$_{0.07}$   under the effect of nematic order and in the absence of magnetic order for the first time.
%At 6.3~K below the superconducting transition temperature, superconducting gap is resolved at the hole pocket around the Z point.
The remarkable anisotropy of the superconducting gap rules out $s$++ pairing purely mediated by orbital fluctuations.
The gap amplitude decreases from Z to $\Gamma$, till it is undetectable at $\Gamma$, which is intriguingly different from that of BaFe$_2$(As$_{1-x}$P$_x$)$_2$ with a nodal ring around Z.
%This may relate to their different stacking periods of FeAs or FeSe layers along the c-axis.
%The superconducting gap is absent at the electron pocket around A, indicating gap nodes or minimal gap size at these momenta.
A 2-fold anisotropy of the superconducting gap is observed at the $\alpha$ hole pocket  around Z, which cannot be understood by current theories unless the effects of nematicity are considered. Our results suggest that in order to  comprehensively understand this unique family of FeSe$_{1-x}$S$_x$,
future theories  should include the effects of nematicity and quantum paramagnetism, where multiple spin fluctuation-mediated pairing channels cooperate.

%based on antiferromagnetic order and simple combinations of different pairing mechanisms.
%The nematic order, short-range antiferromagnetism, quantum paramagnetic ground state, and their effect on the superconductivity should all be taken into account while developing the theories of Fe-based superconductors.
%The intimate interplay between the superconducting pairing and nematic state in FeSe should to be taken into account. These results put strong constraints on the theories of superconductivity in Fe-based superconductors.

\textit{Acknowledgements:} We gratefully acknowledge Moritz Hoesh, Pavel Dudin and Timur Kim  for the experimental assistance at Diamond Light Source,  and Yan Zhang for helpful discussions and sharing his unpublished data of independent measurement, and Jiangping Hu, Dung-Hai Lee, Andrey Chubukov, Rafael Fernandes, and Jian Kang for helpful discussions. This work is supported in part by the National Science Foundation of China, National Basic Research Program of China (973 Program) under the grant No. 2012CB921402, and the Science and Technology Committee of Shanghai under the grant No. 15YF1401000 and 15ZR1402900.

%FWHM%
%orbital in fourth point
%acknowledgement%
%anisotropic fitting%
%fit alpha, too large at M?%
%FeSeS
%Funds

\end{document}